\documentclass[11pt]{article} 
\newcommand{\be}{\begin{equation}}
\newcommand{\ee}{\end{equation}}
\newcommand{\bea}{\begin{eqnarray}}
\newcommand{\eea}{\end{eqnarray}}
\newcommand{\sn}{{\rm sn}}
\newcommand{\ds}{{\rm ds}}
\newcommand{\cs}{{\rm cs}}

\newcommand{\dn}{{\rm dn}}
\newcommand{\cn}{{\rm cn}}
\newcommand{\sech}{{\rm sech}}

\begin{document}
\vspace{.5in} 
\begin{center} 
{\LARGE{\bf Linear Superposition for a Large Number of  Nonlinear 
Equations}}
\end{center} 

\vspace{.3in}
\begin{center} 
{\LARGE{\bf Avinash Khare}} \\ 
{Raja Ramanna Fellow} \\
{Indian Institute of Science Education and Research (IISER), \\
Pune, India 411021}
\end{center} 

\begin{center} 
{\LARGE{\bf Avadh Saxena}} \\ 
{Theoretical Division and Center for Nonlinear Studies, Los
Alamos National Laboratory, Los Alamos, NM 87545, USA}
\end{center} 

\vspace{.9in}
{\bf {Abstract:}}  

We demonstrate a kind of linear superposition for a large number of
nonlinear equations, both continuum and discrete. In particular, 
we show that whenever a nonlinear equation admits solutions in terms of  
Jacobi elliptic functions $\cn(x,m)$ and $\dn(x,m)$, then it also admits 
solutions in terms of their sum as well as difference, i.e. 
$\dn(x,m) \pm \sqrt{m}\, \cn(x,m)$. Further, we also show that whenever a 
nonlinear equation admits a solution in terms of $\dn^2(x,m)$, it also has  
solutions in terms of $\dn^2(x,m) \pm \sqrt{m}\, \cn(x,m)\, \dn(x,m)$ even 
though $\cn(x,m)\, \dn(x,m)$ is not a solution of that nonlinear equation. 
Finally, we  obtain similar superposed solutions in coupled theories. \\ 

\newpage
  
{\bf Introduction:} 
Linear superposition principle is one of the hallmarks of linear theories
which does not hold good in nonlinear theories because of the nonlinear
term. For example, even if two solutions are known for a nonlinear theory, their
superposition is in general not a solution of the nonlinear theory. The
purpose of this letter is to point out a kind of superposition 
for a large number of nonlinear equations. In particular, there are several 
nonlinear equations, both discrete and continuum, which are known to admit 
exact periodic solutions in terms of Jacobi elliptic functions $\cn(x,m)$ 
as well as $\dn(x,m)$, where $m$ denotes the modulus of the elliptic function
\cite{as}. Many of these solutions have found application in several areas of 
physics \cite{1,2}. In particular, we have examined a large number of nonlinear 
equations, both continuum and discrete, which admit both $\cn(x,m)$ and 
$\dn(x,m)$ solutions and find that in all these cases even 
$\dn(x,m) \pm \sqrt{m}\,\cn(x,m)$ is also an exact solution.
We have also examined a number of {\it coupled} field theories and have 
obtained superposed coupled solutions of the form $\dn \pm \sqrt{m} \cn$ 
in both the fields.

Further, we have also examined a number of continuum field theories which 
admit $\dn^2(x,m)$ as a solution and find that such theories also admit 
$\dn^2(x,m) \pm \sqrt{m}\, \cn(x,m)\,\dn(x,m)$ as solutions. While this cannot
be treated as a linear superposition of two solutions (since $\cn(x,m) \dn(x,m)$ 
is not a solution of these models), we find it rather remarkable that such solutions 
exist, without exception, in a number of continuum field theories (including 
several coupled models) that we have looked at so far.

In this letter we only discuss a few selected examples, several
other examples will be given elsewhere in a longer version.  To begin 
with, we discuss one continuum (quadratic-cubic nonlinear Schr\"odinger 
equation or QCNLS) and one  discrete model (saturated discrete nonlinear 
Schr\"odinger equation or DNLS) both of which admit $\cn(x,m)$ as well as 
$\dn(x,m)$ as solutions and show that these models also admit $\dn(x,m) \pm 
\sqrt{m}\, \cn(x,m)$ as solutions. Note that stable kink \cite{qcnls1} and chaotic 
soliton solutions under nonlinearity management \cite{qcnls2} are known for 
the QCNLS equation which arises in such diverse physical systems as nonlinear 
optics, chemical kinetics, matter-radiation interactions and mathematical ecology 
\cite{qcnls1, qcnls2}.  The DNLS equation arises in the context of optical fibre 
communication \cite{of}.  We then discuss a coupled NLS-MKdV model and show 
that it has $\dn(x,m) \pm \sqrt{m}\, \cn(x,m)$ type coupled solutions in both the 
fields. Here MKdV refers to the modified KdV equation \cite{1}. 

Subsequently, we discuss the Korteweg-de Vries (KdV) equation which is known 
to admit $\dn^2(x,m)$ as a solution \cite{1,2} and show that this model also admits 
$\dn^2(x,m) \pm \sqrt{m}\, \cn(x,m)\, \dn(x,m)$ as solutions even though $\cn(x,m)\, 
\dn(x,m)$ is not a solution of the KdV equation. We also show that the coupled 
NLS-MKdV model, not only admits $\dn(x,m) \pm \sqrt{m}\cn(x,m)$ but also 
$\dn^2(x,m)\pm \sqrt{m} \cn(x,m) \dn(x,m)$ type solutions in both the fields.
Further, we show that a coupled NLS-KdV model has superposed solutions of the 
form $\dn(x,m) \pm \sqrt{m}\, \cn(x,m)$ 
- $\dn^2(x,m) \pm \sqrt{m}\, \cn(x,m)\, \dn(x,m)$.  
Finally, we briefly mention possible reasons why such a linear superposition 
works in several nonlinear theories.

{\bf $\dn \pm \sqrt{m}\, \cn$ as Exact Solutions}

We first discuss the quadratic-cubic nonlinear Schr\"odinger equation (QCNLSE) 
and then the discrete NLS equation (DNLSE) and show the existence of such 
superposed solutions in both the cases.

{\bf QCNLSE}: We start with the quadratic-cubic nonlinear Schr\"odinger 
equation 
\be\label{1}
iu_t+u_{xx}+g_1|u|u + g_2|u|^2 u  =0\,.
\ee
One of the exact moving periodic solutions to this equation is 
\be\label{2}
u = \big (A \dn[\beta(x-vt+\delta_1),m]+B \big )\exp[-i(\omega t-kx+\delta)]\,,
\ee
provided
\be\label{3}
g_2 A^2 = 2\beta^2\,,~~g_2 B^2 = (2-m) \beta^2\,,~~g_1 = -3Bg_2\,,~~
~~\omega = k^2 +2(2-m) \beta^2\,,~~v=2k\,.
\ee
Here $\delta,\delta_1$ are two arbitrary constants.

Similarly, another exact moving periodic solution to the QCNLSE, Eq. (\ref{1}), is
\be\label{4}
u = \big (A \sqrt{m} \cn[\beta(x-vt+\delta_1),m]+B) 
\exp[-i(\omega t-kx+\delta)] , 
\ee
provided
\be\label{5}
g_2 A^2 = 2\beta^2\,,~~g_2 B^2 = (2m-1)\beta^2\,,~~g_1 = -3Bg_2
~~\omega = k^2 +2(2m-1)\beta^2\,,~~v=2k\,.
\ee
Notice that the $\cn(x,m)$ solution only exists if $1/2 < m \le 1$.  We now show that, 
remarkably, even though $\cn(x,m)$ solution does not exist if $m \le 1/2$, a linear 
superposition of $\cn(x,m)$ and $\dn(x,m)$ is still an exact solution over the 
entire range $0 < m \le 1$, i.e.
\be\label{6}
u = \bigg ( \frac{A}{2} \dn[\beta(x-vt+\delta_1),m]
+ \frac{D}{2} \sqrt{m}\, \cn[\beta(x-vt+\delta_1),m]+B \bigg )
\exp[-i(\omega t-kx+\delta_2)]\,,
\ee
is an exact solution to the QCNLSE, Eq. (\ref{1}), provided
\bea\label{7a}
&&D = \pm A\,,~~g_2 A^2 = 2 \beta^2\,,~~g_2 B^2 = (1+m)/2 \beta^2\,, 
\nonumber \\
&&g_1 = -3Bg_2\,,~~\omega =k^2 + (1+m)\beta^2\,,~~v=2k\,.
\eea

Here $\cn(x,m)$ and $\sn(x,m)$ are periodic functions with period $4K(m)$, 
$\dn(x,m)$ is a periodic function with period $2K(m)$, with
$K(m)$ being the complete elliptic integral of the first kind \cite{as}. 
It is worth noting that the frequency $\omega$ of the three solutions 
(i.e. $\cn, \dn$ and $\dn \pm \sqrt{m}\, \cn$) is different except at $m=1$. 
We thus have two new periodic solutions of QCNLSE depending on if
$D=A$ or $D=-A$. 

Few remarks are in order here which are in fact valid 
for all the models (both continuum and discrete) that admit such solutions.

\begin{enumerate}

\item Both the solutions $\dn +\sqrt{m}\, \cn$
and $\dn-\sqrt{m}\, \cn$ exist for the same values of the parameters. 

\item In the limit $m=1$, the three solutions $\dn$, $\cn$ as well as 
$\dn+\sqrt{m}\,\cn$ go over to the well known pulse (i.e. $\sech$) solution. 

\item In all the continuum models admitting such solutions,
the factors of $2-m$ or $2m-1$ which appear in the $\dn$ and $\cn$ 
solutions, get replaced by the factor of $(1+m)/2$ in the 
$\dn \pm \sqrt{m}\,\cn$ solutions. 

\item On the other hand, in the discrete theories admitting such solutions
(see below), 
the factor of $\dn(\beta,m)$ or $\cn(\beta,m)$ appearing in $\dn$ and $\cn$
solutions, gets replaced by a factor of $[\dn(\beta,m)+\cn(\beta,m)]/2$
in the $\dn \pm \sqrt{m}\,\cn$ solutions. 

\item In view of the above two points, either the frequency $\omega$ or 
the velocity $v$ (or could be even both) are different for $\dn,\cn$ and 
$\dn \pm \sqrt{m}\, \cn$ solutions except at $m=1$.

\end{enumerate}

{\bf Saturated Discrete NLS Equation:}
We now consider a discrete nonlinear equation which admits both $\dn$ and 
$\cn$ solutions and show that it also has superposed solutions 
$\dn \pm \sqrt{m}\, \cn$. In particular, we consider saturated DNLS equation
which has received great attention in the context of optical fibre 
communication \cite{of}
\be\label{7}
idu_n/dt +[u_{n+1}+u_{n-1}]+ \frac{\nu \mid u_n \mid^2}
{1+ |u_n|^2}u_n = 0\,. 
\ee
One well known periodic solution to this equation is \cite{ak4}
\be\label{8}
u_n = A \dn[\beta(n+\delta_1),m] e^{-i(\omega t +\delta)}\,,
\ee
provided
\be\label{9}
A^2 \cs^2(\beta,m) =1\,,~~\beta = \frac{2K(m)}{N_P}\,,~~
\omega = -\nu = -2\frac{\dn(\beta,m)}{\cn^2(\beta,m)}\,.
\ee

The other solution is
\be\label{10}
u_n = A\sqrt{m} \cn[\beta(n+\delta_1),m] e^{-i(\omega t +\delta)}\,,
\ee
provided
\be\label{11}
A^2 \ds^2(\beta,m) =1\,,~~\beta=\frac{4K(m)}{N_p}\,,~~
\omega = -\nu = - 2 \frac{\cn(\beta,m)}{\dn^2(\beta,m)}\,.
\ee

Remarkably, even a linear superposition of the two is also an exact solution
to the saturable DNLS Eq. (\ref{7}), i.e. it is easily shown using the recently
derived identities for the Jacobi elliptic functions \cite{ak3} that
\be\label{12}
u_n = \bigg (\frac{A}{2} \dn[\beta(n+\delta_1),m]+
\frac{B}{2}\sqrt{m} \cn[\beta(n+\delta_1),m] \bigg ) 
e^{-i(\omega t +\delta)}\,,
\ee
is also an exact solution to Eq. (\ref{7}) provided
\be\label{13}
B = \pm A\,,~~A^2 [\cs(\beta,m)+ \ds(\beta,m)]^2 =4\,,
~~\beta=\frac{4K(m)}{N_p}\,,~~
\omega = - \nu = - \frac{4}{\dn(\beta,m)+\cn(\beta,m)}\,.
\ee
Here $N_p$ is the spatial period of the system, cs$(\beta,m)$=cn$(\beta,m)$/sn$(\beta,m)$ 
and ds$(\beta,m)$=dn$(\beta,m)$/sn$(\beta,m)$.  
As remarked earlier, observe that if we replace $\dn(\beta,m)$ and 
$\cn(\beta,m)$ by $[\dn(\beta,m)+\cn\beta,m)]/2$ in relations (\ref{9})
and (\ref{11}), we recover relations (\ref{13}). Further, the frequency
$\omega$ of the three solutions $\dn, \cn$ and $\dn \pm \sqrt{m}\, \cn$ is 
different except at $m=1$.

{\bf Coupled NLS-MKdV Model:}
We now show that the coupled NLS-MKdV model admits 
$\dn \pm \sqrt{m} \cn$ type solution in both the fields.

In this case the field equations are given by 
\bea\label{a1}
&&iu_t+u_{xx}+g|u|^2 u+ \alpha u v^2=0 , \nonumber \\
&&v_t+v_{xxx} +6v^2 v_{x}+\gamma v(|u|^2)_{x} =0\,.  
\eea
These coupled equations admit four solutions with $u$ being either
$\cn$ or $\dn$ (multiplied by an exponential) and similarly $v$ can be
either $\cn$ or $\dn$. For example, one of the solutions is
\bea\label{a2}
&&u(x,t)=A\exp[-i(\omega t-kx+\delta)] \dn[\beta(x-ct+\delta_1),m]\,,
\nonumber \\
&&v(x,t)=B \sqrt{m} \cn[\beta(x-ct+\delta_1),m]\,,
\eea
provided
\be\label{a3}
c=2k=(2m-1)\beta^2\,,~~\omega=k^2-(2-m)\beta^2+(1-m)\alpha B^2\,,
\ee
\be\label{ax}
gA^2+\alpha B^2 = 2\beta^2\,,~~\gamma A^2+3B^2 = 3\beta^2\,.
\ee

Remarkably, even a linear superposition
\bea\label{a7}
&&u(x,t)=\frac{1}{2}\exp[-i(\omega t-kx+\delta)] 
\bigg (A\dn[\beta (x-ct+\delta_1),m] \nonumber \\
&&+D\sqrt{m}\cn[\beta (x-ct+\delta_1),m] 
\bigg )\,, \nonumber \\
&&v(x,t)= \frac{1}{2} \bigg (B\dn[\beta (x-ct+\delta_1),m] 
+F\sqrt{m}\cn[\beta (x-ct+\delta_1),m] \bigg )\,,
\eea
is an exact solution of coupled  Eqs. (\ref{a1}) provided 
\be\label{a9}
c=2k=(1+m)\beta^2/2\,,~~\omega=k(k-2)\,,~~D=\pm A\,,~~F=\pm B\,,
\ee
while $A,B$ are given by Eq. (\ref{ax}). Note that the signs of $D = \pm A$
and $F = \pm B$ are correlated. 

{\bf $\dn^2 \pm \sqrt{m} \cn\, \dn$ superposed Solutions}

We now discuss one example of a continuum field theory which admits $\dn^2$
as a solution and show that the same model also admits 
$\dn^2 \pm \sqrt{m}\, \cn\, \dn$ as 
solutions, even though $\cn\, \dn$ is not a solution of the model. 

{\bf KdV Equation:}
It is well known that one of the exact solutions to the 
KdV equation \cite{1} 
\be\label{14}
u_t+u_{xxx}+gu u_{x} =0\,,
\ee
is
\be\label{15}
u = A \dn^2[\beta(x-vt+\delta_1),m]\,,
\ee
provided
\be\label{16}
gA = 12 \beta^2\,,~~v= 4(2-m) \beta^2\,.
\ee

Remarkably, even a linear superposition, i.e.
\be\label{17}
u = \frac{1}{2} \bigg (A\dn^2[\beta(x-vt+\delta_1),m]+
B \sqrt{m}\, \cn[\beta(x-vt+\delta_1),m]\, 
\dn[\beta(x-vt+\delta_1),m] \bigg )\,,
\ee
is an exact solution to the KdV Eq. (\ref{14}) provided
\be\label{18}
B = \pm A\,,~~gA = 12 \beta^2\,,~~v= (5-m)\beta^2\,.
\ee
We thus have two new periodic solutions of KdV Eq. (\ref{14}) 
depending on if $B=A$ or $B=-A$.  

Several remarks are in order here which are in fact valid for all the 
solutions of the form $\dn^2 \pm \sqrt{m} \cn\, \dn$.

\begin{enumerate}

\item We find that all the models which admit $\dn^2$ as a solution, 
also admit  $\dn^2 \pm \sqrt{m}\, \cn\, \dn$ as a solution even though 
$\sqrt{m} \cn\, \dn$ is not a solution of these models and that both of the
new solutions exist for the same values of the parameters.

\item In the limit $m=1$, the two solutions $\dn^2$ and  
$\dn^2+\sqrt{m}\, \cn\, \dn$ go over to the well known pulse (i.e. $\sech^2$) 
solution while $\dn^2-\sqrt{m}\, \cn\, \dn$ 
solution goes over to the vacuum solution $u=0$. 

\item The factors of $2-m$ and $\sqrt{1-m+m^2}$ 
which appear in the $\dn^2$ solution, get
replaced by the factor of $(5-m)/4$ and $\sqrt{1+14m+m^2}/4$, respectively,  
in the $\dn^2 \pm \sqrt{m}\, \cn\, \dn$ solutions (see below). As a 
result the velocity $v$ or frequency $\omega$ (or even both) for 
the solutions $\dn^2$ and $\dn^2 \pm\, \sqrt{m} \cn\, \dn$ are different 
(except at $m=1$). 

\end{enumerate}

We now show that the same coupled NLS-MKdV model as given by Eq. (\ref{a1}), 
also admits $\dn^2$ as well as $\dn^2 \pm \sqrt{m}\cn \dn$ type solutions 
in both the fields.

It is easily shown that the coupled system as given by Eq. (\ref{a1})
admits an exact solution
\bea\label{2.28}
&&u = \bigg (A \dn^2[\beta(x-ct+\delta_1),m]+F \bigg ) 
\exp[-i(\omega t-kx+\delta)]\,,
 \nonumber \\
&&v = B \dn^2[\beta(x-ct+\delta_1),m]+D\,,
\eea
provided
\bea\label{2.29}
&&\alpha =3/2\,,~~\gamma A^2 = -3B^2\,,~~\gamma=2g <0\,,~~(z-y)B^2=2\beta^2\,,
\nonumber \\
&&c=2k=4[(2-m)+3z]\beta^2\,,~~\omega = k^2-[4(2-m)+9y+3z]\beta^2\,, 
\nonumber \\
&&z= \frac{D}{B}\,,~~y=\frac{F}{A}=  \frac{[-(2-m) \pm \sqrt{1-m+m^2}]}{3}\,.
\eea
Note that $A\dn^2+B$ is not a solution of either the NLS or MKdV uncoupled 
field equations even though the coupled system admits such a solution.

Remarkably, even a linear superposition, i.e.
\bea\label{2.30}
&&u = \bigg ( F+\frac{A}{2} \dn^2[\beta(x-ct+\delta_1),m] \nonumber \\
&&+ \frac{G}{2} \sqrt{m} \cn[\beta(x-ct+\delta_1),m] 
\dn[\beta(x-vt+\delta_1),m] \bigg )
\exp[-i(\omega t-kx+\delta)]\,, \nonumber \\
&&v = D+\frac{B}{2} \dn^2[\beta(x-ct+\delta_1),m] \nonumber \\
&&+ \frac{H}{2} \sqrt{m} \cn[\beta(x-ct+\delta_1),m] 
\dn[\beta(x-ct+\delta_1),m]\,,
\eea
is an exact solution to Eq. (\ref{a1}) provided 
\bea\label{2.32}
&&G = \pm A\,,~~H = \pm B\,,~~c=2k=[5-m+12z]\beta^2\,, \nonumber \\
&&\omega = k^2-[(5-m)+9y+3z]\beta^2\,, 
~~y=\frac{F}{A}=  \frac{[-(5-m) \pm \sqrt{1+14m+m^2}]}{12}\,,
\eea
while rest of the relations are exactly the same as those given by 
Eq. (\ref{2.29}). Note that the signs of $G = \pm A$ and $H = \pm B$
are correlated. It is worth reminding once again that neither $A\dn^2+B$
nor $A\sqrt{m} \cn \dn$ is an exact solution of either NLS or MKdV field
equations.

{\bf Coupled NLS-KdV Fields:}
Finally, let us consider the following coupled KdV-NLS field equations
\bea\label{6.18}
&&iu_t+u_{xx}+g|u|^2 u+ \alpha u v=0,  \nonumber \\
&&v_t+v_{xxx} +6vv_{x}+\gamma v(|u|^2)_{x} =0\,, 
\eea
where $u$ and $v$ are the NLS and KdV fields, respectively. 
These equations admit coupled solutions of the form $\dn$ - $\dn^2$
and $\cn$ - $\dn^2$. For example, it is easily checked that 
\bea\label{6.19}
&&u(x,t)=A\exp[-i(\omega t-kx+\delta_1)] \dn[\beta(x-ct+\delta),m]\,,
\nonumber \\
&&v(x,t)=B \dn^2[\beta(x-ct+\delta_1),m]+D\,,
\eea
is an exact solution to the coupled Eq. (\ref{6.18}) provided
\be\label{6.20a}
gA^2 +\alpha B = 2 \beta^2\,,~~\gamma A^2 + 6B = 12 \beta^2\,,
\ee
\be\label{6.21}
c=2k=4[2-m+3z]\beta^2\,,~~\omega= k^2-(2-m)\beta^2 -\alpha z B\,,~~
z=\frac{D}{B}\,.
\ee

Remarkably, the same model also admits interesting superposed solutions 
of the form $\dn \pm \sqrt{m} \cn-\dn^2 \pm \sqrt{m} \cn\dn$.
In particular, it is easily checked that 
\bea\label{6.28}
&&u(x,t)=\frac{1}{2}\exp[-i(\omega t-kx+\delta)] 
\bigg (A\dn[\beta (x-ct+\delta_1),m]
\nonumber \\
&&+H\sqrt{m}\cn[\beta (x-ct+\delta_1),m] \bigg )\,, \nonumber \\
&&v(x,t)=\frac{B}{2} \dn^2[\beta (x-ct+\delta_1),m] \nonumber \\
&&+\frac{F}{2} \sqrt{m} \cn[\beta (x-ct+\delta_1),m] 
\dn[\beta (x-ct+\delta_1),m]+D\,,
\eea
is an exact solution to the coupled field equations (\ref{6.18}) provided
\bea\label{6.30}
&&z=\frac{D}{B}\,,~~
H=\pm A\,,~~F= \pm B\,,~~c=2k=[(5-m)+12z]\beta^2\,, \nonumber \\
&&\omega= k^2-\frac{(1+m)\beta^2}{2}-\frac{\alpha}{4}[(1-m)+4z] B\,,
\eea
and further $A,B$ are again given by Eq. (\ref{6.20a}).

{\bf Conclusion.} 
In this letter we have shown that a kind of linear superposition
holds good in the case of quadratic-cubic NLSE and saturated  discrete NLS 
equations in the sense that these models not only admit $\dn(x,m)$,
$\cn(x,m)$ but even $\dn(x,m) \pm \sqrt{m} \cn(x,m)$ type superposed solutions. 
In fact, by now we have examined a large number of both 
discrete and continuum nonlinear equations such as NLS, MKdV, 
mixed KdV-MKdV system, $\lambda\, \phi^4$ field theory, Ablowitz-Ladik 
equation, saturable discrete NLSE, discrete $\lambda\, \phi^4$ field theories, 
among others, and in all these cases we have obtained such superposed 
solutions.  Furthermore, even in several coupled theories such as NLS-MKdV  
we have obtained $\dn(x,m) \pm \sqrt{m}\, \cn(x,m)$  type coupled solution
in both the fields.  

Further, we have shown that for KdV equation, which is known to admit
a periodic solution in terms of $\dn^2(x,m)$, also admits a solution of the form 
$\dn^2(x,m) \pm \sqrt{m} \cn(x,m) \dn(x,m)$. By now, we have examined a 
number of nonlinear equations like quadratic NLSE and in all these cases we 
have obtained such solutions. Further, in several coupled theories like 
NLS-MKdV, quadratic NLS-KdV, and others  too we have obtained $\dn^2(x,m) 
\pm \sqrt{m} \cn(x,m) \dn(x,m)$ type solutions. We have also looked at the 
NLS-KdV coupled system and obtained mixed solutions of the form $\dn(x,m) \pm 
\sqrt{m}\, \cn(x,m)$ - $\dn^2(x,m) \pm \sqrt{m}\, \cn(x,m)\, \dn(x,m)$.  In addition, 
we have examined several other coupled equations such as MKdV-KdV, 
NLS-quadratic NLS, MKdV-quadratic NLS, among others, and there too we 
have obtained such superposed solutions. 

Many of the nonlinear equations that we have looked at have found wide 
application in several interesting physical situations \cite{as,1,2,of,ak4}  
including fibre optics, chemical kinetics and mathematical ecology.  It would 
be worthwhile to enquire if the new solutions that we have obtained have 
some physical relevance, e.g. in optical fibre communication and related 
contexts \cite{of}.  As a first step in that direction, it is important to examine 
the (linear and nonlinear) stability of these newly found solutions in the 
various models.  

We would like to reemphasize that what we have obtained is only a kind of 
linear superposition and not the full superposition as obtained in the linear
theories. However, we find it remarkable that in spite of the nonlinear terms, 
even a kind of linear superposition holds good in these nonlinear models.

What could be the possible reason why such a linear superposition is possible? 
We surmise mathematically that the main reason is that $\cn(x,m)$ and 
$\dn(x,m)$ functions are quite similar and both of them as well as their 
derivatives are identical at $m=1$. This is in contrast to the $\sn(x,m)$ 
elliptic function which is different from both $\cn(x,m)$ and $\dn(x,m)$ 
functions at any value of $m$ and that is why $\cn(x,m) \pm \sn(x,m)$ 
superposition does not seem to work. We have checked it in the various 
examples mentioned above. Perhaps there is a deeper physical reason to all 
these findings which needs to be explored. \\ 

This work was supported in part by the U.S. Department of Energy.

\end{document}